\documentclass{ws-rv9x6}
\usepackage{subfigure}     
\usepackage{ws-rv-van}     
\usepackage{amsmath}
\usepackage{bm}
\usepackage{graphicx}
\makeindex

\begin{document}

\newcommand{\bra}[1]{\left\langle #1 \right|}
\newcommand{\ket}[1]{\left| #1 \right\rangle}
\newcommand{\adag}{a^{\dagger}}
\newcommand{\del}{\partial}
\newcommand{\Hhat}{\hat{H}}
\newcommand{\Nhat}{\hat{N}}
\newcommand{\Qhat}{\hat{Q}}
\newcommand{\Phat}{\hat{P}}
\newcommand{\Ghat}{\hat{G}}
\newcommand{\That}{\hat{\Theta}}
\newcommand{\Nt}{\tilde{N}}
\newcommand{\Hc}{{\cal H}}
\def \cJ{{ \cal J}}
\renewcommand{\d}{\partial}
\newcommand{\dbeta}{\frac{\partial}{\partial\beta}}
\newcommand{\dgamma}{\frac{\partial}{\partial\gamma}}
\newcommand{\ddbeta}{\frac{{\partial}^2}{\partial{\beta}^2}}
\newcommand{\ddgamma}{\frac{{\partial}^2}{\partial{\gamma}^2}}

\chapter[BCS-pairing and nuclear vibrations]
{BCS-pairing and nuclear vibrations}\label{ra_ch1}

\author[K. Matsuyanagi, N. Hinohara, K. Sato]
{Kenichi Matsuyanagi,$^{1,2}$ Nobuo Hinohara$^{3}$ and Koichi Sato$^{1}$
}

\address{$^{1}$ RIKEN Nishina Center, Wako 351-0198, Japan\\
$^{2}$ Yukawa Institute for Theoretical Physics, Kyoto University,  
Kyoto 606-8502, Japan\\
$^{3}$ Department of Physics and Astronomy, University of North Carolina, 
Chapel Hill, North Carolina 27599-3255, USA\\
matsuyanagi@riken.jp
}

\begin{abstract}
On the basis of time-dependent mean-field picture, 
we discuss the nature of the low-frequency quadrupole vibrations 
from small-amplitude to large-amplitude regimes
as representatives of surface shape vibrations of 
a superfluid droplet (nucleus). 
We consider full five-dimensional quadrupole dynamics 
including three-dimensional rotations restoring the broken symmetries 
as well as axially symmetric and asymmetric shape fluctuations. 
We show that the intimate connections between the 
BCS-pairing and collective vibrations reveal through 
the inertial masses governing their collective kinetic energies. 
\end{abstract}

\body

\section{Introduction}
In almost all even-even nuclei consisting of even number 
of protons and neutrons, 
aside from the doubly magic nuclei of the spherical shell model, 
the first excited states possess angular momentum two and positive parity 
($I^\pi=2^+$). 
Their excitation energies are much lower than the energy gap $2\Delta$ 
characterizing nuclei with superfluidity
(see Fig.~4 in the contribution of Bertsch to this volume),
and  very large electric quadrupole ($E2$)  transition strengths 
(in comparison with those of single-particle transitions) 
between these first excited $2^+$ states and the ground states 
have been systematically observed.  
These experimental data clearly indicate that they are collective 
excitations of the superfluid system. 
\cite{ber94,bri05}
They are genuine quantum vibrations essentially different in character 
from surface oscillations of a classical liquid drop,    
that is, superfluidity and shell structure of nuclei 
play indispensable roles for their emergence.  
Unfortunately, this point is quite insufficiently described in several textbooks 
on nuclear physics. 

In a nucleus whose mean field breaks the spherical symmetry but 
conserves the axial symmetry, its first excited $2^+$ state can be   
interpreted as a uniformly rotating state, provided that 
the amplitudes of quantum shape fluctuations are smaller than 
the magnitude of equilibrium deformation.  
Nuclei exhibiting very small ratios of the excitation energy 
to the energy gap, $E(2^+)/2\Delta$,  
(less than about 0.1) belong to this category
(see Fig.~4 in the contribution of Bertsch to this volume). 
The rotational moment of inertia evaluated from $E(2^+)$ 
turned out to be about half of the rigid-body value. 
This was one of the most clear evidences leading to 
the recognition that their ground states are in superfluid phase. 
Large portion of nuclei exhibiting regular rotational spectra 
have the prolate shape. 
Origin of the asymmetry between the prolate and oblate shapes  
is an interesting fundamental problem still under study.\cite{ari12}      

The first excited $2^+$ states other than the rotational states have 
been regarded as quadrupole vibrations around the spherical shape. 
Their frequencies are low and decrease 
as the numbers of neutrons and protons increasingly
deviate from the magic numbers of the spherical shell model. 
Eventually, they turn into the rotational $2^+$ states discussed above. 
Thus, low-frequency quadrupole vibrations may be regarded 
as soft modes of the quantum phase transitions breaking 
the spherical symmetry of the mean field.  
In a finite quantum system like nuclei, however, 
this phase transition takes place gradually 
as a function of nucleon number,  
and there is a wide region of nuclei whose  
low-energy excitation spectra exhibit 
characteristics intermediate between 
the vibrational and the rotational patterns.     
The softer the mean field toward the quadrupole deformation, 
the larger the amplitude and 
the stronger the nonlinearity of the vibration.  

In this Chapter, we discuss mainly 
the low-frequency (slow) quadrupole vibrations   
rather than summing up the diversity of nuclear vibrational phenomena.  
The reason is not only because they dominate in low-lying spectra 
but also because they represent most typically 
the intimate connection between the BCS-pairing 
and the emergence of collective vibrational modes in nuclei.  
Many ideas developed here are applicable also 
to low-frequency octupole ($3^-$) vibrations.   
We here restrict ourselves to the time-dependent mean-field approach,  
because it provides a clear correspondence between 
the quantum and classical aspects of the surface shape vibrations.  
Furthermore, this approach enables us to microscopically derive 
the collective coordinates and momenta on the basis of 
the time-dependent variational principle.   
We shall show that the inertial masses determining the collective 
kinetic energies of the low-frequency quadrupole modes 
clearly reveal their character 
as surface shape vibrations of a superfluid droplet (nucleus). 

We shall start from small-amplitude vibrations around the spherical 
equilibrium shape and then go to large-amplitude regime 
where we need to consider full five-dimensional (5D) quadrupole dynamics 
including three-dimensional rotations restoring the broken symmetries 
as well as axially symmetric and asymmetric shape fluctuations. 
Through this Chapter, we would like to stress that construction of  
microscopic quantum theory of large-amplitude collective motion (LACM) 
is one of the most challenging open subjects in nuclear structure physics. 
Nowadays, the dimension of nuclear collective vibrational phenomena 
awaiting applications of such a microscopic quantum theory is 
enormously increasing covering wide regions from low to highly excited states, 
from small to large angular momenta, 
and from the proton-drip line to the neutron-drip line. 

\section{Collective motion as moving self-consistent mean field}

\subsection{Small-amplitude regime}

Let us consider even-even nuclei 
whose ground states consist of correlated nucleon pairs 
occupying time-reversal conjugate single-particle states.  
The Hartree-Fock-Bogoliubov (HFB) method is a generalized mean-field theory 
treating the formation of the HF mean field 
and the nucleon pair condensate in a self-consistent manner\cite{ben03}, 
and yields the concept of quasiparticles as 
single-particle excitation modes in the presence of the pair condensate. 

As is well  known, Bohr and Mottelson opened the way to  
a unified understanding of single-particle and collective motions  
of nuclei by introducing the concept of moving self-consistent mean field.
\cite{boh76,mot76,boh75}
The time-dependent extension of the HFB mean field, 
called the time-dependent HFB (TDHFB) theory,  
is suitable to formulate their ideas.\cite{bel65,bar68} 
The TDHFB state vector $|\phi(t)\rangle$ 
can be written in a form of generalized coherent state: 
\begin{equation}
|\phi(t)\rangle = e^{i{\hat G}(t)}|\phi(t=0)\rangle = e^{i{\hat G}(t)}|\phi_0\rangle, 
\end{equation}
\begin{equation}
i{\hat G}(t)= \sum_{(ij)} (g_{ij}(t) a_i^{\dag}a_j^\dag - g_{ij}^*(t) a_ja_i) ,
\end{equation}
where the HFB ground state $|\phi_0\rangle$ 
is a vacuum for quasiparticles $(a_i^{\dag}, a_j)$ , 
\begin{equation}
a_i|\phi_0\rangle = 0, 
\end{equation}
with the suffix $i$ distinguishing different quasiparticle states. 
The functions $g_{ij}(t)$ in the one-body operator ${\hat G}(t)$ 
is determined by the time-dependent variational principle
\begin{equation}
\delta\langle\phi(t) \vert
i\frac{\partial}{\partial t} - H \vert \phi(t) \rangle=0.
\end{equation}
For small-amplitude vibrations around a HFB equilibrium point, 
one can make a linear approximation to the TDHFB equations 
and obtain the quasiparticle random phase approximation (QRPA) 
which is a starting point of microscopic theory of collective motion.
\cite{bar60,mar60} 
Expanding Eq.~(4) as a power series of ${\hat G}(t)$ 
and taking only the linear order, we obtain
\begin{equation}
\delta\langle\phi_0 \vert
[ H, i{\hat G} ] + \frac{\partial {\hat G}}{\partial t} 
\vert \phi_0 \rangle=0.
\end{equation}
Writing ${\hat G}(t)$ in terms of the creation and annihilation operator 
$(\Gamma^\dag,\Gamma)$ of the excitation mode as
\begin{equation}
i{\hat G}(t) = \eta(t) \Gamma - \eta^*(t) \Gamma^\dag,
~~~~~\eta(t)=\eta e^{-i\omega t}, 
\end{equation}
we obtain the QRPA equation which determines 
the microscopic structure of  $(\Gamma^\dag,\Gamma)$ 
as a coherent superposition of many two-quasiparticle excitations. 
Alternatively, we can write ${\hat G}(t)$ 
in terms of the collective coordinate and momentum operators 
(${\hat Q}, {\hat P}$) 
and their classical counterparts ($q(t), p(t)$) as 
\begin{equation}
{\hat G}(t) = p(t) {\hat Q} - q(t) {\hat P}  
\end{equation}
and obtain the QRPA equation,
\begin{eqnarray}
~[~{\hat H}, ~{\hat Q}~ ] &=& -i {\hat P}/D, ~~\\ 
~[~{\hat H}, ~{\hat P}~ ] &=&  i C {\hat Q},   
\end{eqnarray}
for (${\hat Q}, {\hat P}$). 
Here $C$, $D$ and $\omega^2 = C/D$ respectively denote 
the stiffness,  the inertial mass and the frequency squared of 
the vibrational mode (with $\hbar=1$). 
For Anderson-Nambu-Goldstone (ANG) modes,
\cite{and58,nam60}
$C$ and $\omega$ are zero but $D$ are positive.  
Note that Eqs.~(8) and (9) can be used 
also for unstable HFB equilibria 
where $C$ is negative and $\omega$  is imaginary. 
For simplicity, we assumed above that there is only a single 
collective mode, but in reality 
${\hat G}(t)$ is written as a sum over many QRPA normal modes. 

The self-consistent mean field of a finite quantum system 
generates a variety of shell structure dependent on its shape, 
and single-particle wave functions possess 
individual characteristics.  
In addition to rich possibilities of spatial structure,  
collective excitations associated with 
the spin-isospin degrees of freedoms of nucleons occur.  
Thus, diversity of collective vibrations emerges.
\cite{har01,bor98}
Even restricting to the $2^+$ surface oscillation,  
there are two modes of different characters.   
One is the low-frequency mode 
generated mainly from two-quasiparticle excitations 
within partly filled major shells (for both protons and neutrons).
The other is the high-frequency mode, called giant quadrupole 
resonance, generated from single-particle excitations across 
two major shells.  
While giant resonances are small amplitude vibrations, 
low-frequency collective modes in open shell nuclei 
exhibit significant nonlinear effects 
and we need to go beyond the QRPA. 
In the QRPA, the quadrupole vibrational modes 
can be regarded as phonons of 5D harmonic oscillator 
and excitation spectra are expected to show a simple pattern: 
e.g., the two-phonon states (double excitations of the $2^+$ quanta) 
will appear as a triplet with  $I^\pi=0^+, 2^+$ and $4^+$. 
Closely examining experimental data, e.g., on their $E2$ transition properties, 
one finds that they often exhibit significant anharmonicities 
even when a candidate of such a triplet is seen.
\cite{gar10}
The vibrational amplitude becomes very large 
in transient situations of the quantum phase transition from spherical to deformed,   
where the spherical mean field is barely stable or  
the spherical symmetry is broken only weakly. 
Many nuclei are situated in such transitional regions. 

\subsection{Quadrupole collective dynamics}

One of the microscopic approaches to treat nonlinear vibrations is 
the boson expansion method, 
where the collective QRPA normal modes at the spherical shape are 
regarded as bosons and nonlinear effects are evaluated 
in terms of a power series expansion with respect to 
the boson creation and annihilation operators. 
This method has been widely used 
for low-energy collective phenomena.
\cite{kle91}.   

In the investigation of low-energy excitation spectra, 
the pairing-plus-quadrupole (P+Q) model
\cite{kis63, bes69}
and its extension
\cite{sak91}
have played a central role. 
This phenomenological effective interaction 
represents the competition between the pairing correlations 
favoring the spherical symmetry 
and the quadrupole (particle-hole) correlations 
leading to the quadrupole deformation of the mean field. 
Combining the P+Q model with the TDHFB theory,   
Belyaev\cite{bel65}, Baranger and Kumar\cite{bar68} 
microscopically derived the 5D quadrupole collective Hamiltonian 
describing the quadrupole vibrations and rotations in a unified manner: 
\begin{equation}
H= T_{\rm vib}+ T_{\rm rot}+V(\beta,\gamma), 
\end{equation}
\begin{equation}
T_{\rm vib}=\frac{1}{2}D_{\beta\beta}(\beta,\gamma)\dot \beta^2+D_{\beta\gamma}(\beta,\gamma)\dot \beta \dot
\gamma+\frac{1}{2}D_{\gamma\gamma}(\beta,\gamma)\dot \gamma^2, 
\end{equation}
\begin{equation}
T_{\rm rot}=\sum_{k}\frac{ I_k^2}{2\cJ_k(\beta,\gamma)}. 
\end{equation}
Here, $T_{\rm vib}$ and $T_{\rm rot}$ denote the kinetic energies of vibrational 
and rotational motions, while $V(\beta,\gamma)$ represents 
the collective potential energy  
defined through the expectation value of an effective interaction 
with respect to the TDHFB state. 
The velocities of the vibrational motion are described in terms of 
the time-derivatives ($\dot{\beta}$, $\dot{\gamma}$) of 
the quadrupole deformation variables ($\beta$, $\gamma$)  
representing the magnitude and the triaxiality of the quadrupole deformation, 
respectively. 
They are defined in terms of the expectation values of the quadrupole moments 
or through a parametrization of the surface shape. 
The three components $I_k$ of the rotational angular momentum 
and the moments of inertia 
$\cJ_k=4\beta^2D_k(\beta,\gamma)\sin^2(\gamma-2\pi k/3)$ 
in the rotational energy $T_{\rm rot}$
are defined with respect to the intrinsic frame of reference;  
that is,  an instantaneous principal-axis frame of the 
time-dependent shape-fluctuating mean field. 

After quantization with the Pauli prescription, the vibrational kinetic energy 
takes the following form:\cite{pro09}
\begin{eqnarray}
\hat T_{\rm vib}=-\frac{1}{2\sqrt{WR}}
\left[ 
\frac{1}{\beta^4} \dbeta \beta^2\sqrt{\frac{R}{W}}
\left( D_{\gamma\gamma}\dbeta-D_{\beta\gamma}\dgamma \right)  
\right.
\nonumber \\
\left.
-\frac{1}{\beta^2\sin 3\gamma} \dgamma \sqrt{\frac{R}{W}}\sin 3\gamma
\left( D_{\beta\gamma}\dbeta-D_{\beta\beta}\dgamma \right)
\right], 
\end{eqnarray}
where
\begin{eqnarray}
W&=&\beta^{-2}\left[
D_{\beta\beta}(\beta,\gamma)D_{\gamma\gamma}(\beta,\gamma)
-D_{\beta\gamma}^2(\beta,\gamma) \right],  \\
R&=&D_1(\beta,\gamma)D_2(\beta,\gamma)D_3(\beta,\gamma). 
\end{eqnarray}
If the functions, $D_{\beta\beta}, D_{\gamma\gamma}/\beta^2$ 
and $D_k$, are replaced with a common constant  
and $D_{\beta\gamma}$ is ignored, 
then Eq.~(13) reduces to a simpler expression used in many papers. 
However, such a drastic approximation is valid only 
for small-amplitude vibrations around the spherical shape.  
In general situations, it is mandatory to take into account 
the $\beta$ and $\gamma$ dependences of the inertial functions 
as well as the $D_{\beta\gamma}$ term. 

In an axially deformed nucleus whose collective potential $V(\beta,\gamma)$
has a deep minimum at $\beta \ne 0$ and
$\gamma=0^\circ$ (prolate shape) or $\gamma=60^\circ$ (oblate shape), 
a regular rotational spectrum appears. 
At the same time, one can identify 
$\beta$ and $\gamma$ bands involving vibrational quanta of 
fluctuations of the $\beta$ and $\gamma$ variables. 
Close examination of their properties,  however, reveals 
significant nonlinear character of the $\gamma$ vibration.
\cite{mat85a}
It has been known that the $\beta$ vibration couples, often strongly,  
with the pairing vibration associated with the fluctuation of the 
pairing gap $\Delta$.  Recent experiments reveal further interesting 
features of the excited $0^+$ states 
\cite{hey11}
upon which we shall touch in Section 3.  

\subsection{Quantum shape fluctuations and collective rotations 
restoring the broken symmetry}

As is well known, the fundamental concept 
underlying the BCS superconductivity is 
`spontaneous symmetry breaking and appearance of collective modes 
restoring the broken symmetry.'
\cite{and58,nam60}
Nuclear rotation can be regarded as a manifestation of this dynamics  
in a finite quantum system; that is, 
it is a collective motion restoring the spherical symmetry 
broken by the self-consistently generated mean field.
\cite{boh76,boh75}
It is important, however, to keep in mind that 
any HFB equilibrium shape inevitably accompanies  
quantum zero-point fluctuations.  
The well-known $I(I+1)$ pattern of rotational spectrum  
will not appear if the fluctuation amplitude is larger 
than the equilibrium value of $\beta \ne 0$.  
Even when the minimum in the collective potential $V(\beta,\gamma)$ 
is deep in the $\beta$ direction, it may be soft with respect to  
the $\gamma$ direction breaking the axial symmetry. 
In the axially symmetric limit, 
the rotational motion about the symmetric axis is absent. 
However, as soon as the axial symmetry is dynamically broken 
by quantum shape fluctuations, all rotational degrees of freedom 
about the three principal axes (of the instantaneous shape) are activated.   
Low energy excitation spectrum in such a situation exhibits 
a feature more complex than the simple rotational pattern. 
It seems that many nuclei belong to this category. 

\subsection{Microscopic theory of LACM}

The TDHFB theory describes the time evolution of 
the superfluid mean field 
without explicitly introducing collective variables.  
To derive the collective Hamiltonian, we have to assume that 
the time evolution is governed by a few collective coordinates and momenta. 
In the work of Baranger and Kumar,
\cite{bar68}
the 5D collective Hamiltonian was  
derived by giving the role of collective coordinates to the quadrupole 
operators.  We note, however, that there are two kinds of  $2^+$ vibration,  
and the high frequency quadrupole giant resonance carries the major part 
(about 90$\%$, see Fig.~5 in the contribution of Bertsch to this volume) 
of the energy-weighted sum-rule value for the quadrupole operator. 
On the other hand, the collective variables are defined in terms of  
the low-frequency $2^+$ QRPA modes  
in the derivation of the 5D collective Hamiltonian by means 
of the boson-expansion method. 
\cite{sak91}    
In the QRPA modes, contributions of the two-quasiparticle excitations 
near the Fermi surface are much larger than those in the quadrupole operators.
Therefore, the two definitions are different significantly. 

Attempts to construct microscopic theory of LACM 
on the basis of the TDHFB mean field 
dates back to the latter half of the seventies
(see Refs.~\refcite{dan00,mat10} for reviews). 
The major challenge was how to extract
the collective submanifold embedded in the TDHFB phase space, 
which is maximally decoupled from other microscopic 
degrees of freedom.\cite{mar80}  
Once such a collective submanifold is extracted, 
we can set up local canonical coordinates on it. 
Such canonical coordinates may be called ``collective coordinates."  
Below we sketch the basic ideas of the LACM theory. 

Let us assume that the time evolution of the TDHFB state  
is determined  by the collective coordinate $q(t)$ and momentum $p(t)$. 
To restore the gauge invariance broken by 
the HFB mean-field approximation for superfluid nuclei,
it is necessary to find a way extending the QRPA procedure 
to non-equilibrium. 
For this purpose, 
we introduce the number fluctuation variable $n(t)$ and 
the gauge angle $\varphi(t)$ conjugate to it and 
write the TDHFB state vector in the following form:  
\begin{eqnarray}
\ket{\phi(q,p,\varphi,n)} &= e^{-i \varphi\Nt}\ket{\phi(q,p,n)},\\
\ket{\phi(q,p,n)} &= e^{ip\Qhat(q) + in\That(q)}\ket{\phi(q)}.
\label{eq:TDHFBstate}
\end{eqnarray}
Here $\ket{\phi(q,p,n)}$ represents an intrinsic state for  
the pairing rotational degree of freedom parametrized by $\varphi$, 
$\ket{\phi(q)}$  a non-equilibrium HFB state,  
$\Nt$ nucleon number fluctuation, and  
$\Qhat(q), \That(q)$ infinitesimal generators.  
We also define an infinitesimal displacement operator $\Phat(q)$ by
\begin{equation}
\ket{\phi(q + \delta q)} = 
e^{-i \delta q \Phat(q)}\ket{\phi(q)}. 
\end{equation}
Microscopic structures of 
$\Qhat(q), \Phat(q), \That(q)$ and $\ket{\phi(q)}$ are determined 
on the basis of the time-dependent variational principle:  
\begin{eqnarray}
\delta \bra{\phi(q,p,\varphi,n)} i\frac{\del}{\del t} - \Hhat \ket{\phi(q,p,\varphi,n)}  = 0,
\label{eq:TDVP}
\end{eqnarray}
where $\Hhat$ is a microscopic many-body Hamiltonian. 
(For simplicity, we assume that there is only a single canonical set of 
collective variables.) 

Let us assume that time variation of the mean field is slow 
(in comparison with the single-particle motion in the mean field), 
and expand $\ket{\phi(q,p,n)}$ in powers of $p$ and $n$. 
Requiring that the time-dependent variational principle be satisfied 
at each order, we obtain the equations determining the
infinitesimal generators, $\Qhat(q), \Phat(q),$ and $\That(q)$, 
which are a generalization of the QRPA about an HFB equilibrium 
to non-equilibrium HFB states. 
Solving these equations in a self-consistent way, 
we obtain a classical collective Hamiltonian written 
in terms of canonical variables,  which can be readily quantized 
and yield the collective Schr\"odinger equation 
for collective wave functions.   
The procedure outlined above has been formulated 
as the adiabatic self-consistent collective coordinate (ASCC) method.
\cite{mat00}
Quite recently, we have developed a practical approximation scheme 
called ``constrained HFB+ local QRPA (LQRPA) method" 
to efficiently carry out such calculations.
\cite{hin10}
Examples of numerical application are presented in Figs.~1 and 2. 
In both cases, we see clear correlations 
between the $\beta$-$\gamma$ dependence  
of the pairing gap $\Delta$ and of the inertial mass $D_{\beta\beta}$; 
that is, $D_{\beta\beta}$ becomes small in the region where $\Delta$ is large. 

\begin{figure}[ht]
\centerline{
  \subfigure[Excitation spectra]
     {\epsfig{figure=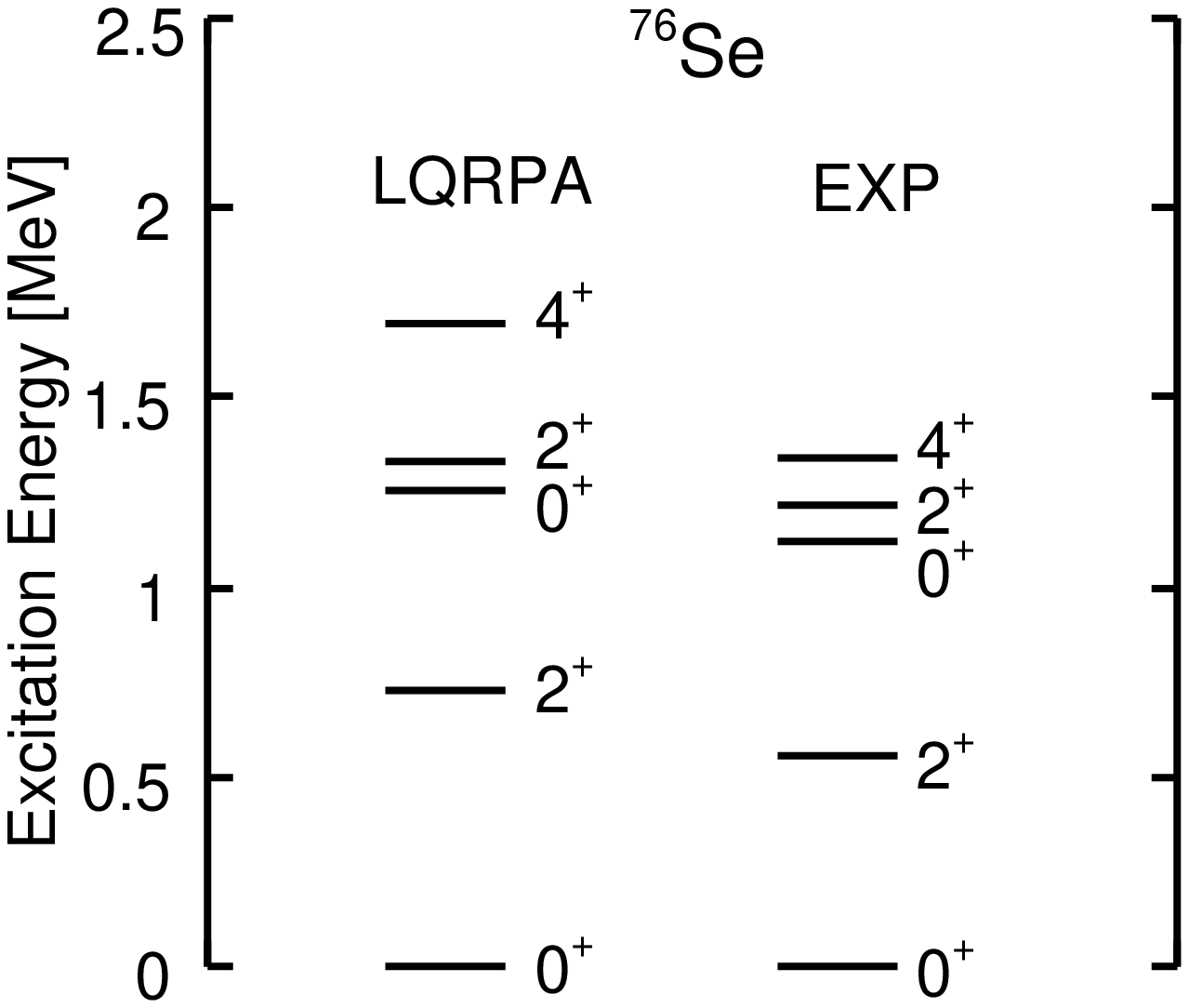,width=2.3in}\label{ra_fig1a}}
  \subfigure[Neutron pairing gap]
     {\epsfig{figure=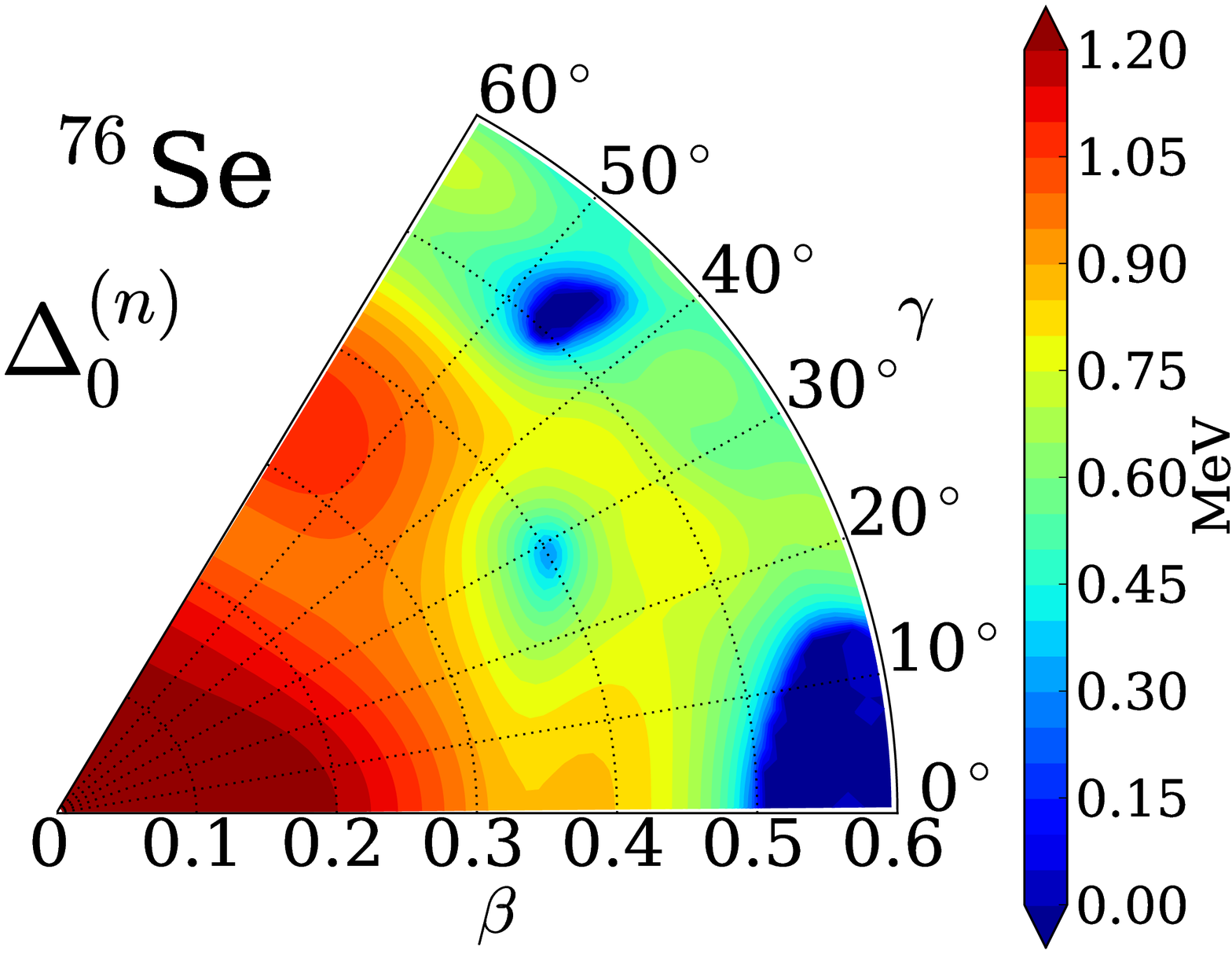,width=2.3in, trim= 10 30 10 30}\label{ra_fig1b}}
}
\vspace{2mm}
\centerline{
  \subfigure[Collective potential] 
     {\epsfig{figure=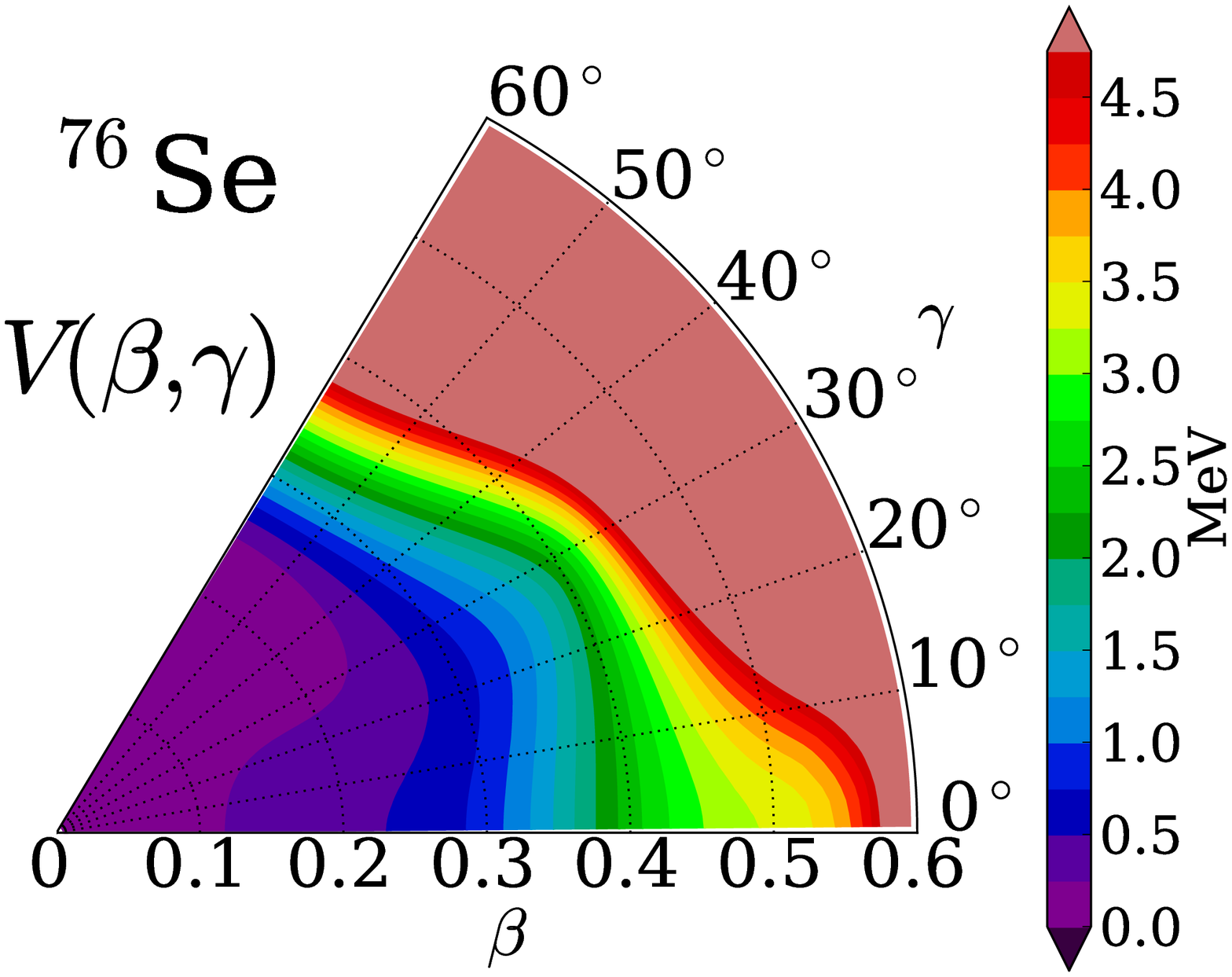,width=2.3in, trim= 10 30 10 30}\label{ra_fig1c}}
  \subfigure[Vibrational mass]
     {\epsfig{figure=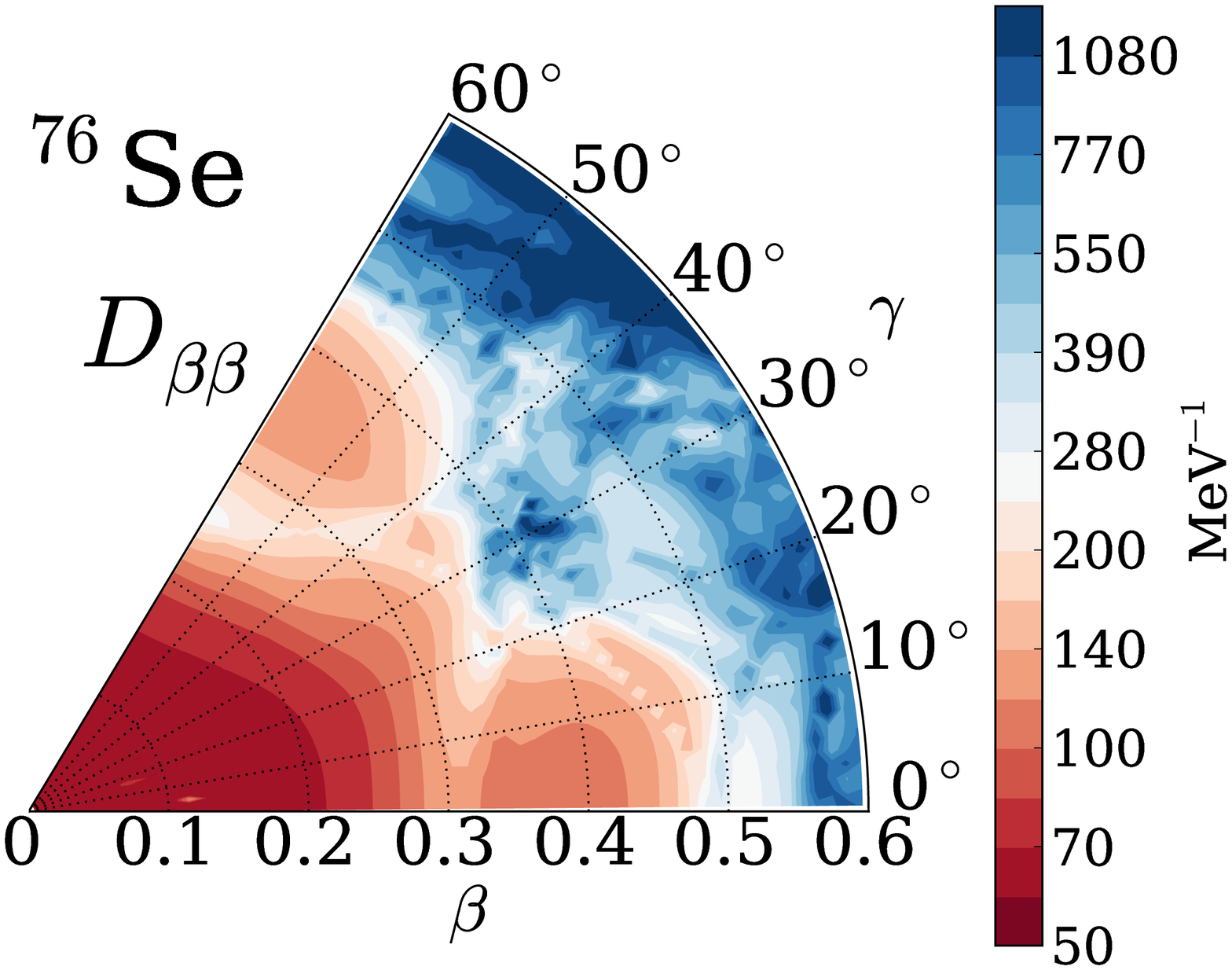,width=2.3in, trim= 10 30 10 30}\label{ra_fig1d}}
}
\caption{Application of the LQRPA method to 
anharmonic quadrupole vibrational excitations in $^{76}$Se. 
Note that the colors are used differently for $\Delta_0^{(n)}$ and $D_{\beta\beta}$. 
}\label{ra_fig1}
\end{figure}

\subsection{Microscopic mechanism of determining the inertial mass}

The reason why the pairing correlation plays a crucial role
in determining the inertia mass of collective motion may be understood 
in the following way.
\cite{bar90}
The single-particle energy levels change following the motion of the mean field 
and encounter a number of level crossings. 
When a level crossing occurs near the Fermi surface, 
the lowest-energy configuration changes. 
Without the pairing, it is not always easy to rearrange the system  
to energetically more favorable configurations.
In the presence of the pairing correlation, in contrast, 
it is easy for nucleon pairs to hop from up-sloping levels to down-sloping levels.   
The easiness/hardness of the configuration rearrangements at the 
level crossings determines the adiabaticity/diabaticity of the collective motion. 
Since the inertia represents a property of the system trying 
to keep a definite configuration, we expect that  
the stronger the pairing, the smaller the inertial mass.  

In this connection, let us note the following fact. 
The nucleon pair in a deformed mean field is not simply 
a monopole ($J=0$) pair but 
a superposition of different angular momenta $J$, 
because the spherical symmetry is broken.  
Especially,  one cannot ignore the quadrupole pairing correlations 
acting among the $J=2$ components. 
For example, when the prolately deformed nucleus develops toward a larger 
value of $\beta$,  single-particle energy levels favoring the prolate shape 
go down while those favoring the oblate shape go up. 
At their level crossing point, the ability of the rearrangement depends 
on the pairing matrix element between the crossing levels. 
The spacial overlap between the prolate-favoring and the oblate-favoring 
single-particle wave functions is smaller than its value at the spherical limit. 
This effect is taken into account by including the quadrupole pairing correlation. 
If this effect is ignored, the inertial mass will be underestimated.
\cite{hin10}
The interaction strengths of the monopole and quadrupole components  
are linked by the requirement of Galilean invariance.
\cite{sak91}

\section{Remarks on some current topics}

\subsection{Shape coexistence and quantum shape fluctuations}

In the situations where two different HFB equilibrium shapes coexist 
in the same energy region, LACM tunneling through 
the potential barrier between the two HFB local minima may take place. 
This is a macroscopic tunneling phenomenon 
where the potential barrier itself is generated as a 
consequence of the dynamics of the self-bound quantum system. 
For instance, two strongly distorted rotational bands built on 
the oblate and prolate shapes have been found in $^{68}$Se, 
which seems to coexist and interact with each other. 
\cite{fis00}
Figure 2 shows an application of the LQRPA method 
to this oblate-prolate shape coexistence/fluctuation phenomenon.
\cite{hin10}
Such phenomena are widely seen in low-energy spectra 
from light to heavy nuclei. 
\cite{hey11}

\begin{figure}[ht]
\centerline{
\epsfig{figure=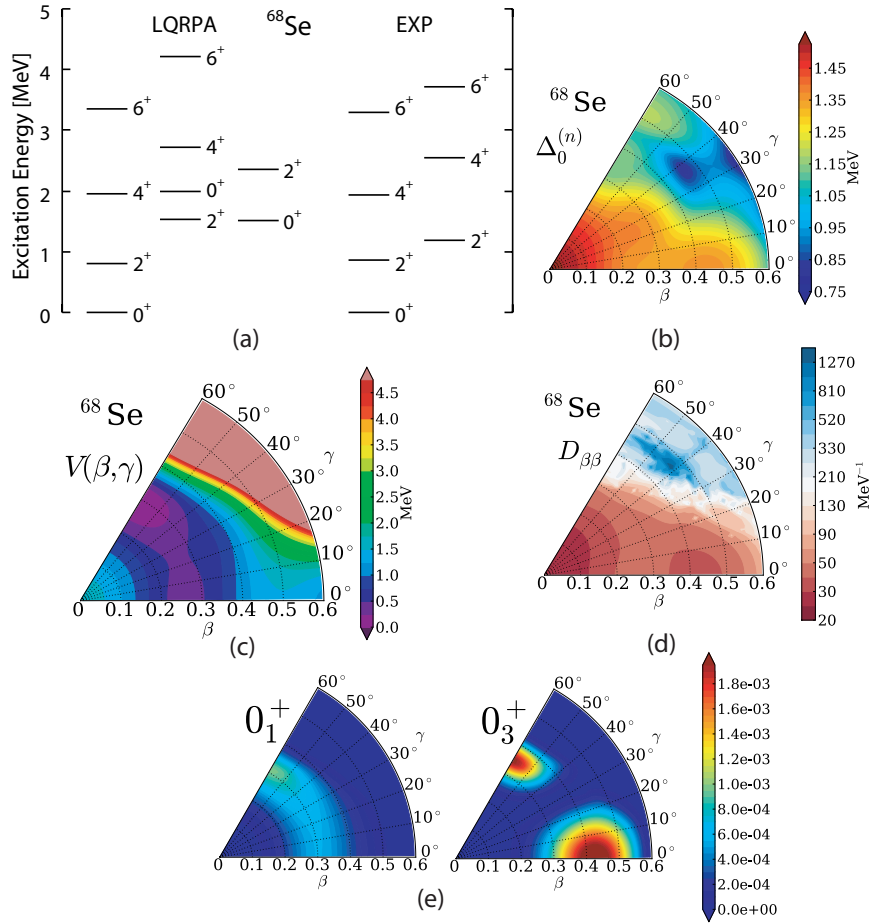,width=\textwidth}
}
\caption{
Application of the LQRPA method\cite{hin10} to 
the oblate-prolate shape coexistence/fluctuation phenomenon in $^{68}$Se. 
In the bottom part, vibrational wave functions squared 
(multiplied by $\beta^4$) 
for the ground $0_1^+$ and the (experimentally unknown) third $0_3^+$ states 
are displayed. The $\beta^4$ factor takes into account 
the major $\beta$ dependence 
of the volume element for the 5D collective Hamiltonian.  
}
\label{ra_fig2}
\end{figure}

One of the recent hot issues related to the shape coexistence/fluctuation 
is to clarify the nature of deformation in neutron-rich nuclei 
around $^{32}$Mg, 
where  two-particle-two-hole configurations of neutrons
across the spherical magic number  $N=20$ play a crucial role.\cite{hey11}
It seems that the pairing and quadrupole correlations 
act coherently in this situation to generate 
a large-amplitude quadrupole shape fluctuations. 
\cite{hin11}

\subsection{Mysterious $0^+$ excited states}

There are only a few nuclei in which the first excited  $0^+$ state 
appears below the first excited $2^+$ state. 
An example is the $0^+$ state of $^{72}$Ge 
which is known from old days but still poorly understood. 
This anomaly occurs in the vicinity of $N=40$ 
where the $g_{9/2}$ shell starts to be partly filled (due to the pairing). 
It has been pointed out
\cite{tak86,wee81}  
that the neutron pairing vibrations strongly couple with 
the quadrupole vibrations there and generates  
such anomalous $0^+$ states. 
It is an open problem whether such $0^+$ excited states are describable 
within the 5D quadrupole dynamics or it is mandatory to extend 
the dimension of the collective submanifold explicitly treating  
the pairing gaps as dynamical variables.     
Closely examining the properties of low-lying excited $0^+$ states 
throughout the nuclear chart, 
one finds that they exhibit features difficult to understand 
within the traditional models of nuclear collective motions.
\cite{hey11}

\subsection{Vibrational modes at high angular momentum} 

Experimental data for low-frequency vibrations near 
the high-spin yrast states (`ground' states for given angular momenta) 
are scarce. 
As the nucleus rotates more rapidly, 
excitations of aligned quasiparticles take place step by step,  
\cite{fra01,sat05}
the shell structure changes with varying mean-field,  
and the pair field may eventually disappear.
\cite{shi89}
Such drastic changes of the mean-field 
and the presence of aligned quasiparticles 
will significantly modify the properties of vibrational motions.   
The presence of low-frequency vibrations itself is not self-evident, 
if we recall that the BCS pairing plays an essential role in the 
emergence of the low-frequency $2^+$ vibrations. 
On the other hand, we could also expect that  
vibrations may compete with rotations in high-spin yrast region,  
because the rotational frequency increases with 
increasing angular momentum and eventually   
become  comparable to vibrational frequencies.        
\cite{boh81}

Discovery of superdeformed bands opened a new perspective 
to the above open question.  
We learned that a new shell structure,   
called superdeformed shell structure, is formed and 
a new type of soft octupole vibrations 
simultaneously breaking the axial symmetry and 
space-reflection symmetry emerge 
in the near yrast regions of rapidly rotating superdeformed nuclei.
\cite{nak96,ros01}
Quite recently, a number of new data suggesting appearance 
of $\gamma$-vibrations (shape fluctuation modes toward triaxial deformation) 
at high spin have been reported.\cite{oll11,pat03} 
Appearance of triaxial deformation at high spin due to the 
weakening of the pairing correlation has been discussed for a long time, 
but it is only recent years that a variety of experimental data 
unambiguously indicating the triaxial deformation has been obtained.  
New rotational modes appearing when the mean field breaks 
the axial symmetry, called wobbling motions, have been discovered. 
\cite{ode01}
It is shown that the aligned quasiparticle plays 
an important role for their emergence. 
\cite{sho09}
Another new type of rotational spectra expected to appear 
in triaxially deformed nuclei under certain conditions is  
the chiral rotation
\cite{fra01} 
and experimental search for the predicted 
chiral doublet bands and its precursor phenomena, 
called chiral vibrations, 
\cite{alm11} 
are now going on. 

\subsection{Vibrational modes near the neutron drip line}

The mean field in unstable nuclei near the neutron drip line possesses  
new features like large neutron-to-proton ratios, 
formation of neutron skins, weak binding of single-particle states 
near the Fermi surface, excitations of neutron pair into the continuum.
\cite{mat10b}
In stable nuclei, overlaps of different single-particle wave functions 
become maximum at the surface and generate a strong coherence 
among quasiparticle excitations.    
In unstable nuclei, weakly bound single-particle wave functions 
significantly extend to the outside of the half-density surface   
and acquire strong individualities.   
It is therefore very interesting to investigate 
how the pairing correlation in such a situation  
acts to generate the collectivity of vibrational modes.   
It is suggested, for instance, in a recent HFB+QRPA calculation 
simultaneously taking into account 
the deformations of the mean field, the pairing correlations 
and the excitations into the continuum,\cite{yos08} 
that a strong coherence of the pairing and shape fluctuations  
may generate collective vibrations unique to weakly bound neutron-rich nuclei.     

\subsection{Concluding remarks}

Quite recently, it becomes possible to carry out 
fully self-consistent QRPA calculations 
on the basis of density functional theory for superfluid nuclei 
and treat low- and high-frequency vibrations 
as well as the ground states in a unified way 
for all nuclei from the proton-drip line to the neutron-drip line. 
\cite{ter08,eba10,yos11a}
Fully self-consistent microscopic calculations 
for large-amplitude vibrations are also initiated. 
\cite{yos11b}
A new era toward understanding vibrational motions of nuclear 
superfluid droplets is opening.

\end{document}